# Surface variation analysis of freeform optical systems over surface frequency bands for prescribed wavefront errors

RUNDONG FAN,[1] SHILI WEI,[1] HUIRU JI,[2] ZHUANG QIAN,[1] HAO TAN,[2] YAN MO,[2] AND DONGLIN MA[1,2,3] *

[1]*School of Optical and Electronic Information, Huazhong University of Science and Technology, Wuhan, Hubei 430074, China*
[2]*MOE Key Laboratory of Fundamental Physical Quantities Measurement & Hubei Key Laboratory of Gravitation and Quantum Physics, PGMF and School of Physics, Huazhong University of Science and Technology, Wuhan 430074, China*
[3]*Shenzhen Huazhong University of Science and Technology, Shenzhen 518057, China*
* *Corresponding author: madonglin@hust.edu.cn*

**Abstract:** The surface errors of freeform surfaces reflect the manufacturing complexities and significantly impact the feasibility of processing designed optical systems. With multiple degrees of freedom, freeform surfaces pose challenges in surface tolerance analysis in the field. Nevertheless, current research has neglected the influence of surface slopes on the directions of ray propagation. A sudden alteration in the surface slope will lead to a corresponding abrupt shift in the wavefront, even when the change in surface sag is minimal. Moreover, within the realm of freeform surface manufacturing, variation in surface slope across different frequency bands may give rise to unique surface variation. Within the context of this study, we propose a tolerance analysis method to analyze surface variation in freeform surfaces considering surface frequency band slopes based on real ray data. This approach utilizes real ray data to rapidly evaluate surface variation within a specified frequency band of surface slopes. Crucially, our proposed method yields the capability to obtain system surface variation with significant wavefront aberration, in contrast to previous methodologies. The feasibility and advantages of this framework are assessed by analyzing a single-mirror system with a single field and an off-axis two-mirror system. We expect to integrate the proposed methodology with freeform surface design and manufacturing, thereby expanding the scope of freeform optics.

## 1. Introduction

In the last 10 years, freeform optics has enabled compact and high-performance imaging systems. Freeform surfaces can be defined as surfaces with no axis of rotational invariance (within or beyond the optical part) [1]. Freeform surfaces find extensive applications in various domains, including beam shaping [2-5], consumer electronics [6-8], infrared systems [9], scanning systems [10], and refractive-diffractive hybrid systems [11], primarily owing to their remarkable performance and integration potential. The rapid advancement of automatic design algorithms for freeform surfaces has given rise to methodologies such as Simultaneous multiple surfaces (SMS) [12,13], Partial differential equations (PDEs) [14,15], Construction-Iteration (CI) [16-18] and neural network [19,20]. This advancement effectively reduces the trial-and-error process for designers, helping them escape local minimums and significantly enhancing the efficiency of freeform surface design. Designers can now accomplish a substantial amount of design work in less time. However, the inherent asymmetry of freeform surfaces presents manufacturing challenges, which, in turn, hinder the attainment of the optical system's intended design performance. The challenge stems from several key factors. Firstly, freeform optical systems are susceptible to surface errors. Many researchers have attempted to decrease this susceptibility by controlling the deflection angles of rays during the design process [21-23].

Secondly, current tolerance theories may not effectively guide the manufacturing of freeform surfaces. Presently, tolerance for freeform surfaces still relies on measured metrics such as peak-to-valley (PV) and root-mean-square (RMS), which are evaluated by Monte Carlo analysis techniques. Nevertheless, the local characteristics of freeform surfaces cannot be adequately described solely through PV or RMS metrics. While some scholars have suggested the use of Gaussian radial basis functions [24] to represent errors within the surface profiles of freeform surfaces, this method further improves the freedom of expression of surface profile errors. By superimposing Gaussian function of different parameters on the surface and performing Monte Carlo analysis, surface tolerances are finally obtained. The expression of Gaussian radial basis function improves the degree of freedom of describing the surface error. However, the method still does not consider the variation for different regions of the surface based on imaging performance.

In recent years, Deng et al. introduced the novel concept of local tolerance for freeform surfaces [25]. Distinct regions of the optical surface exhibit different effects on the wavefront aberration, leading to different tolerance ranges for these distinct regions. These ranges are denoted as local tolerances. Initially, the upper and lower boundaries for the wavefront aberration associated with sampled rays are defined. Geometric relationships are then utilized to superimpose the optical path difference of these sampled rays into the optical surface along the normal direction at their intersection point. Then a point cloud envelope is systematically computed across the full field of view (FOV) to determine local tolerances and a subsequent Monte Carlo analysis is performed. The effectiveness of local tolerances for freeform surfaces was demonstrated. Local tolerance offers significant potential in desensitization design and freeform surface manufacturing. Tolerance analysis models that do not account for variation in the surface normal due to adjustments in ray intersection positions can be effectively approximated in optical systems characterized by minimal sag variation. In the case of freeform optical systems with minor wavefront aberrations, this approach proves highly effective in determining local tolerances. Nevertheless, the manufacture of freeform surfaces is an ongoing process. When optical surfaces are perturbed, it not only affects the corresponding surface sag but also shifts the position where rays intersect the surface, resulting in variation in the normal vector at those points. In reflective systems, changes in the normal direction can lead to doubled offsets in outgoing rays. In freeform optical systems characterized by significant wavefront aberrations, the influence of surface normal vector fields on optical properties becomes more pronounced. As a result, this method has the potential to introduce errors that could result in inaccurate tolerance outcomes.

Numerous techniques have been employed to manufacture high-performance freeform surfaces, including methodologies such as single-point diamond processing [26,27], computer-controlled surface forming [28,29], and ion beam polishing [30,31], among others. It must be recognized that different manufacturing approaches result in diverse frequency surface errors, and these different frequency errors can have varying impacts on optical system performance. Lower-frequency errors primarily influence image shape and focusing position, while medium and high-frequency errors will introduce aberrations and give rise to structured image artifacts [32-34]. Conducting tolerance analysis for freeform surfaces across various frequency bands allows for a more accurate representation of real-world freeform surface manufacturing. However, developing a tolerance analysis method for surface errors originating from different manufacturing methods remains a significant challenge in tolerance analysis.

This paper presents a tolerance analysis method for analyzing surface variation in freeform optical systems, which includes variation of both surface sag and surface slope related to error frequencies based on real ray data. Our approach accounts for the impact of changes in the surface normal field on imaging performance during the computation process. Within a predefined surface error frequency range, this method enables the rapid obtainment of surface variation spectrum for freeform surfaces. We formulate a nonlinear equation system with constraints, incorporating Fermat's principle, constraints on wavefront aberrations, and constraints related to surface slope related to frequency. The solution to this system of nonlinear equations is obtained through numerical methods. Subsequently, a nested iteration process is employed to reduce operation errors and improve computation efficiency. In practice, this algorithm translates into notable advantages, as demonstrated by the ability to obtain system surface variation with significant wavefront aberration efficiently. Illustrative examples include the single-field single-mirror system and a two-mirror off-axis system.

The main contributions of this work can be summarized as follows. Firstly, the paper presents a new tolerance analysis method for evaluating surface variation centered around the surface frequency band slope based on real ray data. Compared with the existing freeform surface variation analysis methods, the proposed method has the advantage of efficiency. Secondly, our proposed method is universal and practical and can tackle the surface variation of freeform surfaces with significant wavefront aberrations by taking the effect of the surface normal field on rays into consideration. Finally, the method enables the derivation of the coefficient of local surface variation directly and thus is suitable for optical manufacturing by providing a reference for surface tolerance analysis connected with actual manufacturing methods. Moreover, the proposed method can hold the potential to usher in future's freeform surface desensitization design.

## 2. Theory

Both surface sag and slope impact the performance of optical systems. In this section, we offer a succinct analysis of the influence of surface slope on the performance of optical systems, establish constrained nonlinear equations based on real ray data, Fermat's principle, slope constraints, and wavefront constraints, and elucidate the numerical processes. Subsequently, a nested iteration is employed to reduce operation errors and improve computational efficiency.

When a ray travels through a perturbed optical system, it experiences both positional shifts due to refraction or reflection and changes in the surface normal orientation. As shown in Fig. 1, we represent the positional displacement of a refracted or reflected ray as $\Delta d$ and the change in the surface normal as $\theta$. In reflective systems, the impact of perturbations on surface slope is a crucial factor that requires careful consideration. To illustrate this point, we analyze a real optical system. Figures. 2 depicts this system as a parabolic reflective system with a single FOV measuring 0.1°. The wavefront aberration PV and RMS values are presented in Figs. 2(a). When we introduce two sets of surface errors simultaneously, each with different slope frequencies, it becomes evident in Figs. 2(b) and (c) that the PV and RMS values of surface errors in Figs. 2(c) are smaller than those in Figs. 2(b). Nevertheless, the wavefront aberration in Figs. 2(c), characterized by more pronounced slope changes, exhibits a significant increase. This phenomenon negatively impacts the system imaging performance.

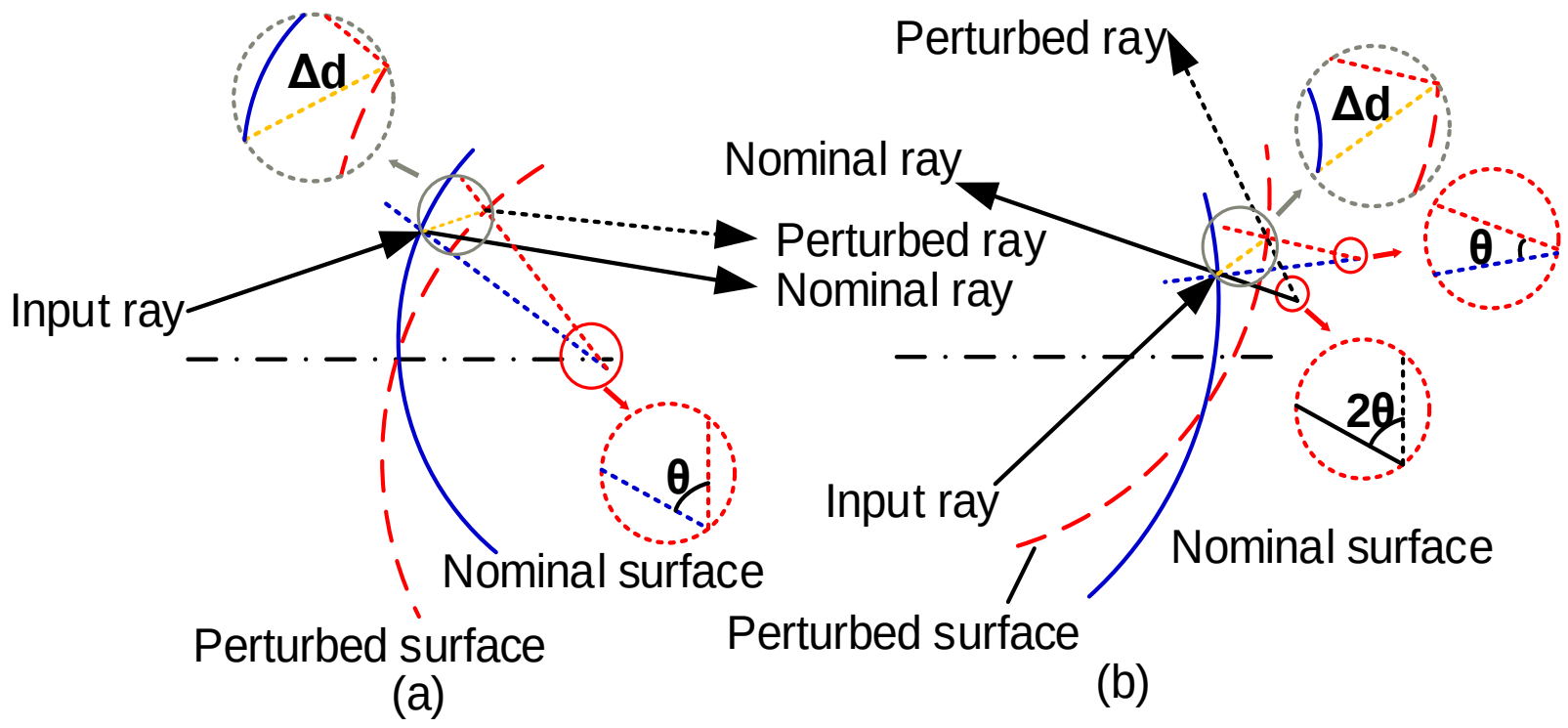

Fig. 1. Slightly disturbed optical surface:(a) refractive optical surface; (b) reflective optical surface.

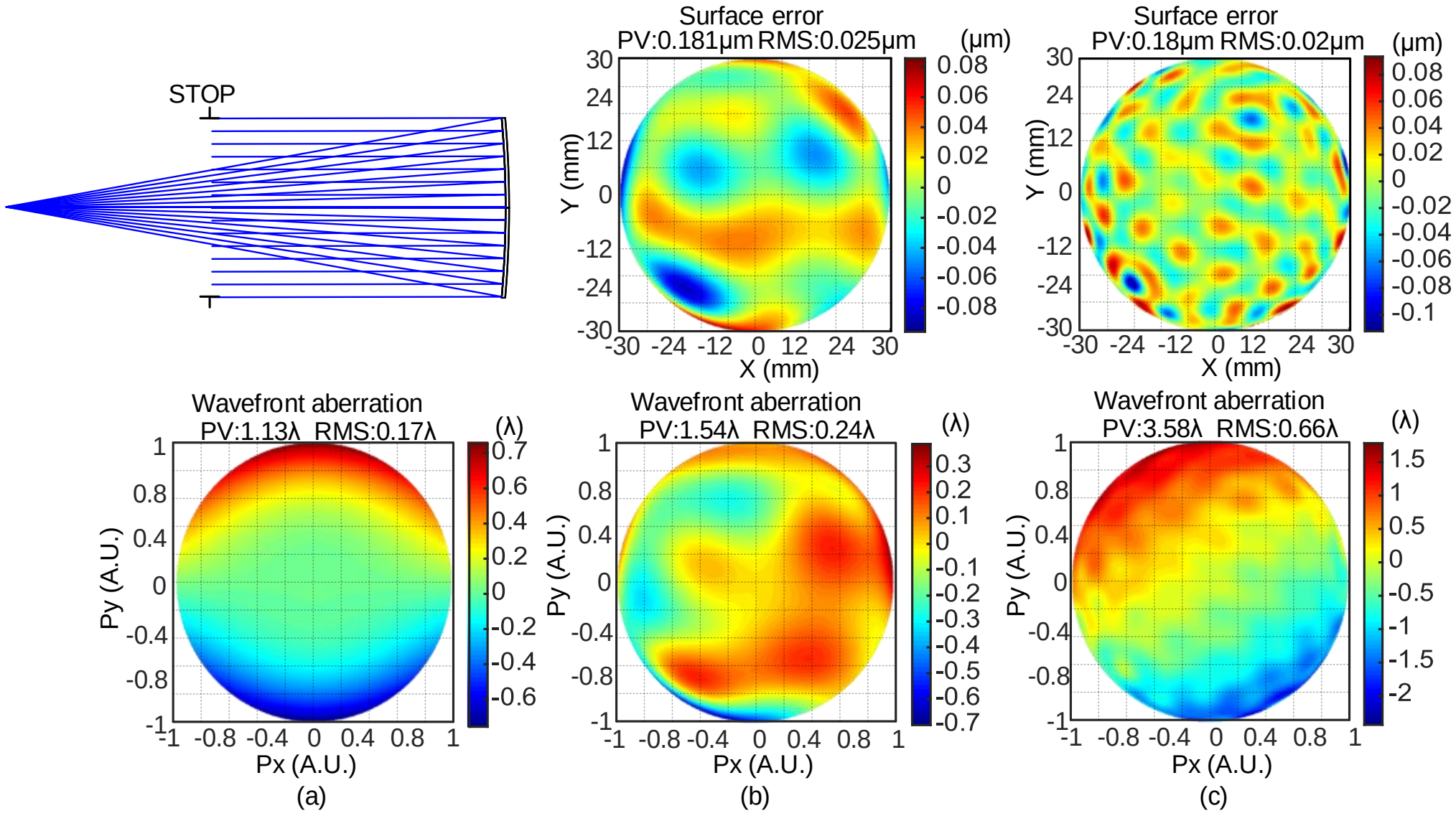

Fig. 2 (a) Single parabolic reflection system and nominal wavefront aberration; (b) relax slopes surface errors and wavefront aberration; (c) serious slope surface errors and wavefront aberration

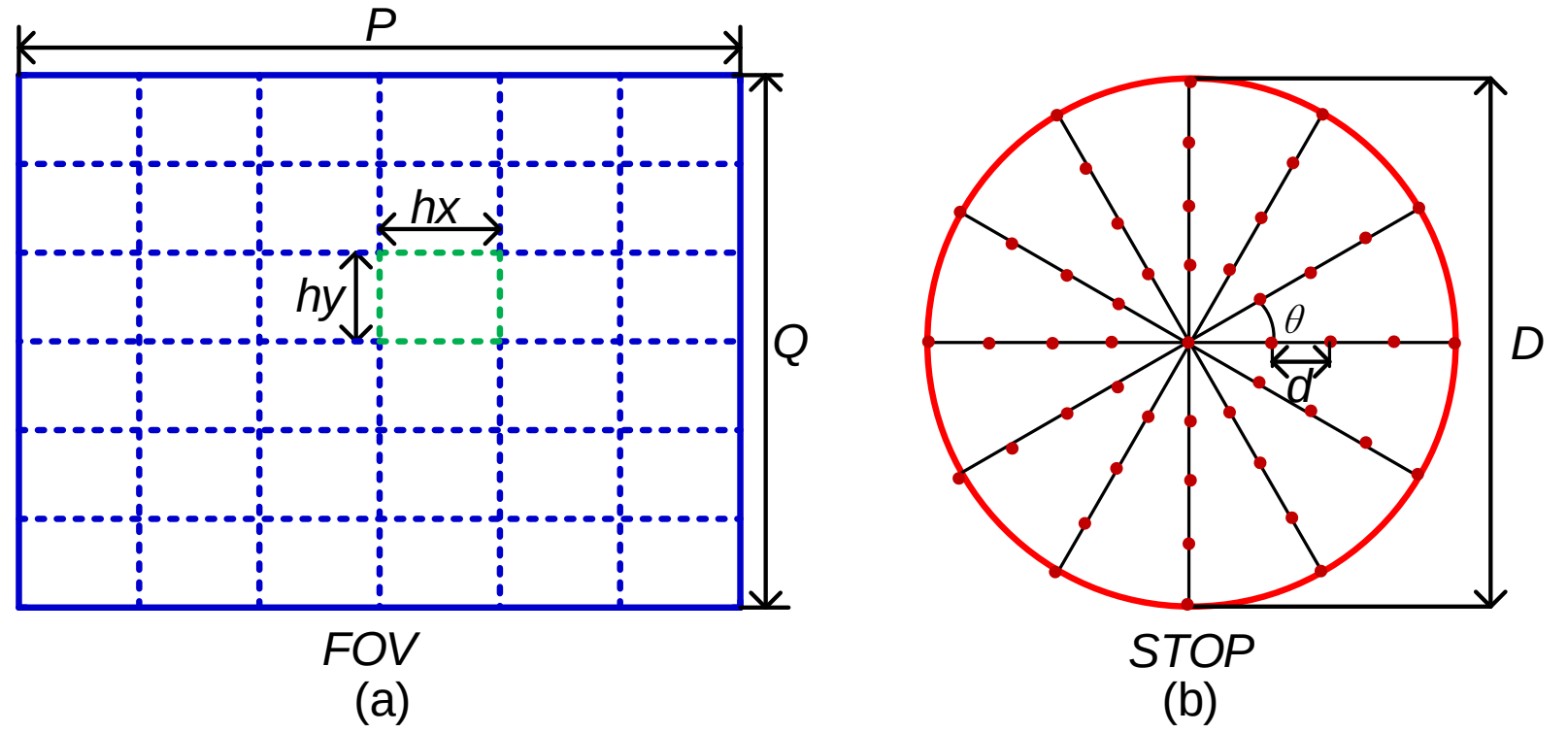

Fig. 3. Characteristic ray sampling: (a) field sampling; (b) stop sampling.

We first sample characteristic rays inside the stop and full FOV. This sampling process is illustrated conceptually in Fig. 3. The grid sampling of the horizontal FOV (referred to as $P$) and the vertical FOV (denoted as $Q$) is illustrated conceptually in Figs. 3(a). The steps in the $x$ and $y$ directions are defined as $hx$ and $hy$. The full FOV is subdivided into equally spaced fields of $F=hx\times hy$. The stop with a size of $D$ is sampled in polar coordinates by specifying the angle step $\theta$ and radial step $d$, which is showcased in Figs. 3(b). The number of samples is determined by $R=(2\pi/\theta)\times(r/d)$.

Firstly, we designate wavefront aberration as the performance metric. To control the optical path length (OPL) of the system, we maintain the wavefront aberration within the expected range. This process results in the determination of upper and lower variation boundaries that correspond to the boundary range of the wavefront aberration, as depicted in Fig. 4. When the wavefront difference experiences a change of $\Delta W$, it leads to a corresponding alteration of $\Delta D$ in the surface sag.

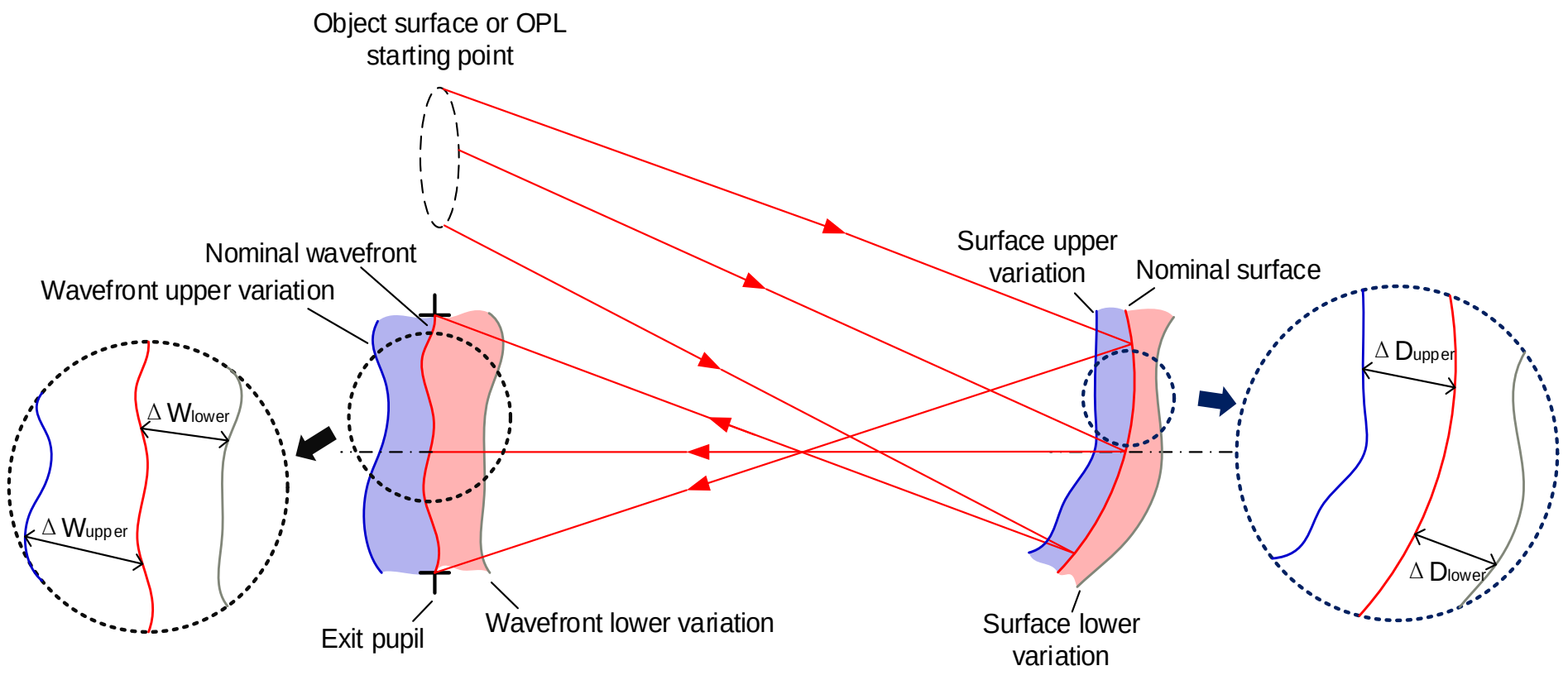

Fig. 4 Wavefront deviation and surface variation diagram

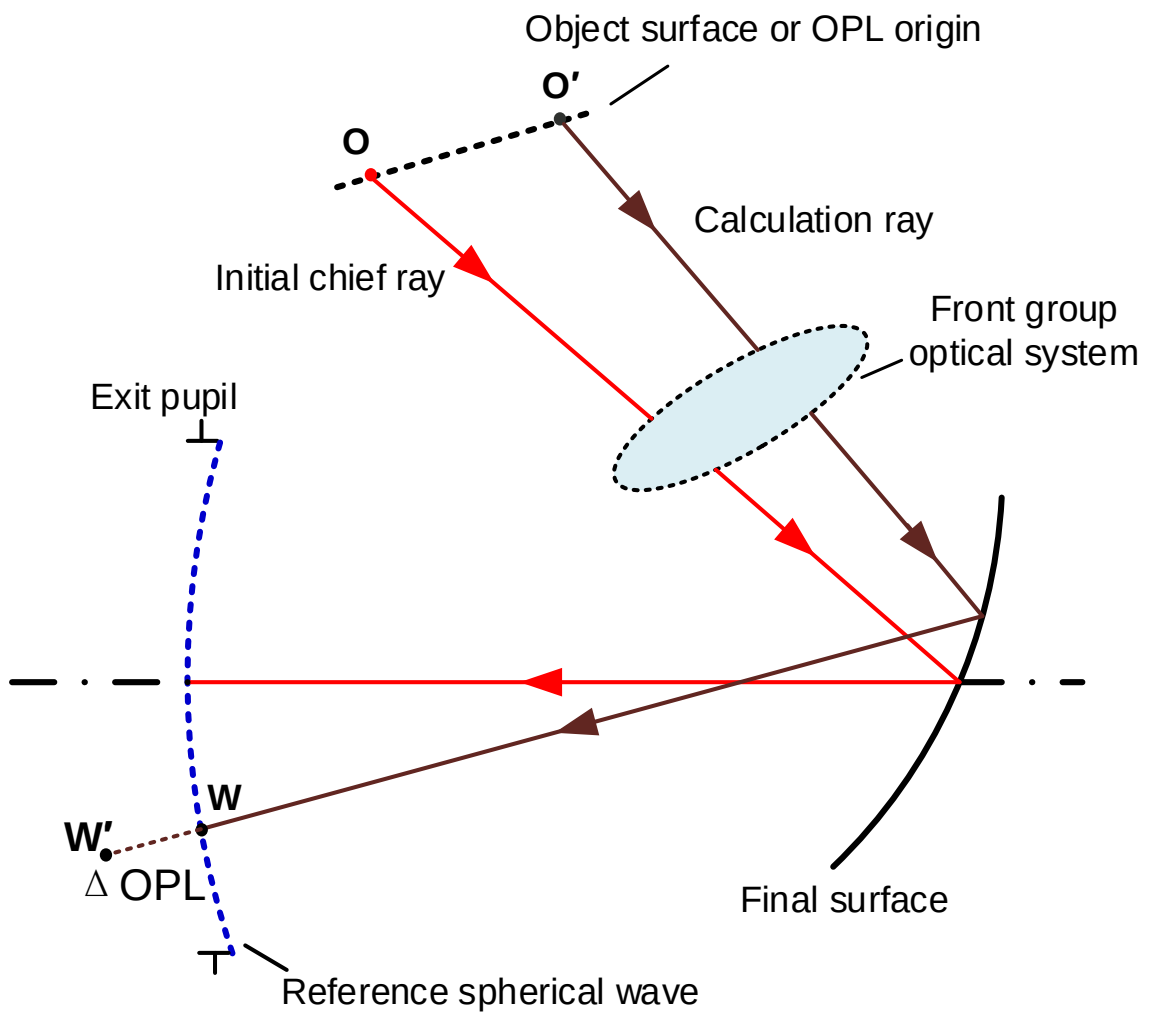

Fig. 5 Schematic of optical path difference in the single field.

When examining a single field, the variation obtaining procedure for the optical surface is depicted in Fig. 5. Starting from the origin of the optical path, the red ray in Fig. 5 represents the initial chief ray within a designated FOV. The total OPL from **O** to **W** by this ray is denoted as $T_{\text{initial}}$, which remains constant throughout the numerical solution process. The brown ray in Fig. 5 represents $j^{\text{th}}$ perturbation ray within the same FOV, characterized by an optical path from **O'** to **W'** denoted as $T_j$. Therefore, the equation for total OPL and OPL constraint can be expressed as:

$$T_j = \sum_{i=0}^{N} n_i L_i \, (i = 0 \ldots N), \tag{1}$$

$$-\frac{W_{\text{Exp}}}{2} \le (T_{\text{initial}} - T_j) \le \frac{W_{\text{Exp}}}{2}, \tag{2}$$

$$(T_{\text{initial}}^{\text{upper}} - T_j^{\text{upper}}) = \frac{W_{\text{Exp}}}{2}, \tag{3}$$

$$(T_{\text{initial}}^{\text{lower}} - T_j^{\text{lower}}) = -\frac{W_{\text{Exp}}}{2}, \tag{4}$$

where from the origin of the optical path $\mathbf{O_n}$, the ray is divided into $N$ optical Spaces by $M$ optical elements, and $n_i$, $L_i$ are the path length of the ray in the $i^{\text{th}}$ optical space and the refractive index of the corresponding space, as shown in Fig.6. The optical space includes the image and object spaces in the lens imaging, with optical elements referring to lenses or mirrors in the system. The $W_{\text{Exp}}$ is wavefront aberration PV threshold. The $T_{\text{initial}}^{\text{upper}}$ and $T_{\text{initial}}^{\text{lower}}$ represent the initial OPL used to obtain upper variation and lower variation. The $T_j^{\text{upper}}$ and $T_j^{\text{lower}}$ represent the $j^{\text{th}}$ OPL used to obtain upper and lower variation values. The upper and lower variation values of the surface variation can be solved by Eq. (3) and Eq. (4).

By linking Eq (3) and Eq (4), the wavefront constraint of full FOV ray can be expressed as:

$$f_1=\sqrt{\sum_{f=1}^{F}\sum_{j=1}^{R}((T_{j,\text{initial}}^{\text{upper/lower}}-T_{j}^{\text{upper/lower}})\pm\frac{W_{\text{Exp}}}{2}))^2 \Big/ (F*R)}, \tag{5}$$

where $F$ and $R$ represent the number of sampled fields and the number of rays sampled in each field.

Slope root-mean-square (SlopeRMS) is an energy-related parameter frequently employed to impose constraints on mirror quality in optical manufacturing [35]. The equation of SlopeRMS is expressed as follows:

$$S=\sqrt{\frac{1}{(M-1)(N-1)}\sum_{i=1}^{M-1}\sum_{j=1}^{N-1}((\frac{Z(x_{i+1,j},y_{i,j})-Z(x_{i,j},y_{i,j})}{\Delta x})^2+(\frac{Z(x_{i,j},y_{i,j+1})-Z(x_{i,j},y_{i,j})}{\Delta y})^2)}, \tag{6}$$

$$Z(x,y)=z_{\text{error}}(x,y), \tag{7}$$

where $M$, $N$ are the total number of sampled pixels in the $x$ and $y$ direction of the surface, and $Z$ is the sag of the surface. The variables $z_{\text{error}}(x,y)$ represent the error sag. The quantities $\Delta x$ and $\Delta y$ correspond to spatial periods in the $x$ and $y$ directions, respectively. Therefore, Eq. (6) represents a function of spatial frequency, allowing for an analysis of SlopeRMS across various frequencies given a specific frequency.

The frequency boundary for the analysis is defined before the operation, and during the derivative process, the surface error slope is constrained to remain within the range of SlopeRMS. This constraint facilitates the surface variation analysis of surface slopes at various spatial frequencies, resulting in surface variation values of different spatial frequencies. The formula for the surface slope constraint is as follows:

$$S_{\text{low}}\le\sqrt{\sum_{f=1}^{F}\sum_{j=1}^{R}(Z_x(x_{f,j},y_{f,j})^2+Z_y(x_{f,j},y_{f,j})^2)\Big/(F*R-1)}\le S_{\text{high}}, \tag{8}$$

$$Z_x(x,y)=\partial z_{\text{error}}(x,y)/\partial x, \tag{9}$$

$$Z_y(x,y)=\partial z_{\text{error}}(x,y)/\partial y, \tag{10}$$

where $S_{\text{low}}$ is low-frequency SlopeRMS, and $S_{\text{high}}$ is high-frequency SlopeRMS. $F$ and $R$ represent the number of fields sampled among full FOV and the number of rays sampled for each field.

Next, we establish the functional relationship between the OPL and the surface point. This is achieved by tracing all specific characteristic rays within the FOV to acquire the real coordinate positions of the rays on the optical surface and their corresponding reference spherical wavefront. As shown in Fig. 6, the analyzed surface is $\mathbf{S}_{\text{k}}$, whose front and rear surfaces are $\mathbf{S}_{\text{k-1}}$ and $\mathbf{S}_{\text{k+1}}$ respectively. $\mathbf{S_{k-1}}$, $\mathbf{S_k}$, $\mathbf{S_{k+1}}$, and $\mathbf{W_n}$ denote the points at which the ray intersects $S_{\text{k-1}}$, $S_{\text{k}}$, $S_{\text{k+1}}$, and the reference spherical wavefront. The formulas for OPL before and after $k^{\text{th}}$ optical surface can be expressed as follows:

$$OPL_k(L_{k-1},L_k)=n_{k-1}L_{k-1}+n_kL_k, \tag{11}$$

where $n_{\text{k-1}}$, $n_{\text{k}}$ are the refractive index of the medium before and after the optical surface. The $L_{\text{k-1}}$, $L_{\text{k}}$ are the length of ray path before and after the optical surface. Taking surface $S_{\text{k}}$ as an example, its calculation method is outlined as follows:

$$L_{k-1}(\mathbf{S}_{k-1}(x_F, y_F, z_F(x_F, y_F)), \mathbf{S}_k(x_P, y_P, Z_p(x_P, y_P))) = \left\| \mathbf{S}_k - \mathbf{S}_{k-1} \right\|_2, \tag{12}$$

$$L_k(\mathbf{S}_{k+1}(x_E, y_E, z_E(x_E, y_E)), \mathbf{S}_k(x_P, y_P, Z_p(x_P, y_P))) = \left\| \mathbf{S}_{k+1} - \mathbf{S}_k \right\|_2, \tag{13}$$

$$Z_p(x.y) = z_{\text{nominal}}(x, y) + z_{\text{error}}(x, y), \tag{14}$$

where $\| \cdot \|_2$ represents second vector norm operation which is handled by the distance formula between two points. The [$x_F$, $y_F$, $z_F(x_F, y_F)$], [$x_E$, $y_E$, $z_E(x_E, y_E)$] are coordinate of point $\mathbf{S_{k-1}}$ and $\mathbf{S_{k+1}}$. [$x_P$, $y_P$, $Z_P(x_P, y_P)$] is coordinated of point $\mathbf{S_k}$. The $Z_P(x_P, y_P)$ is the surface sag of surface $S_k$ derived by Eq. (14). The variables $z_{\text{nominal}}(x,y)$ represent the nominal surface sag.

The OPL of the geometric optical system adheres to Fermat's principle. Fermat's principle thus implies that $DL_{s,x} = \partial(OPL_s(L_{s-1}, L_s))/\partial x = 0$ and $DL_{s,y} = \partial(OPL_s(L_{s-1}, L_s))/\partial y = 0$ [36]. Following similar arguments, we can derive two sets of differential equations for all defined OPL pairwise from object to reference spherical wavefront. The Fermat's principle can be expressed as:

$$\begin{cases} DL_x^s = \partial(OPL_s(L_{s-1}, L_s))/\partial x = 0 (s = 1...M) \\ DL_y^s = \partial(OPL_s(L_{s-1}, L_s))/\partial y = 0 (s = 1...M) \end{cases}, \tag{15}$$

where an optical system consisting of $M$ surface is thus fully described by $M$ differential equations $DL_x^s$ and $M$ differential equations $DL_y^s$ for $s$=1…$M$. Given that tolerance analysis is not universally performed on every surface within certain optical systems, such as the windshield in HUD optical systems. Consequently, variation in surface sag are confined to the surfaces undergoing tolerance analysis, while the OPL for surfaces not undergoing such analysis remains unaltered. Thus, the corresponding equation $DL_x^s$ and $DL_y^s$ for these surfaces equates to 0.

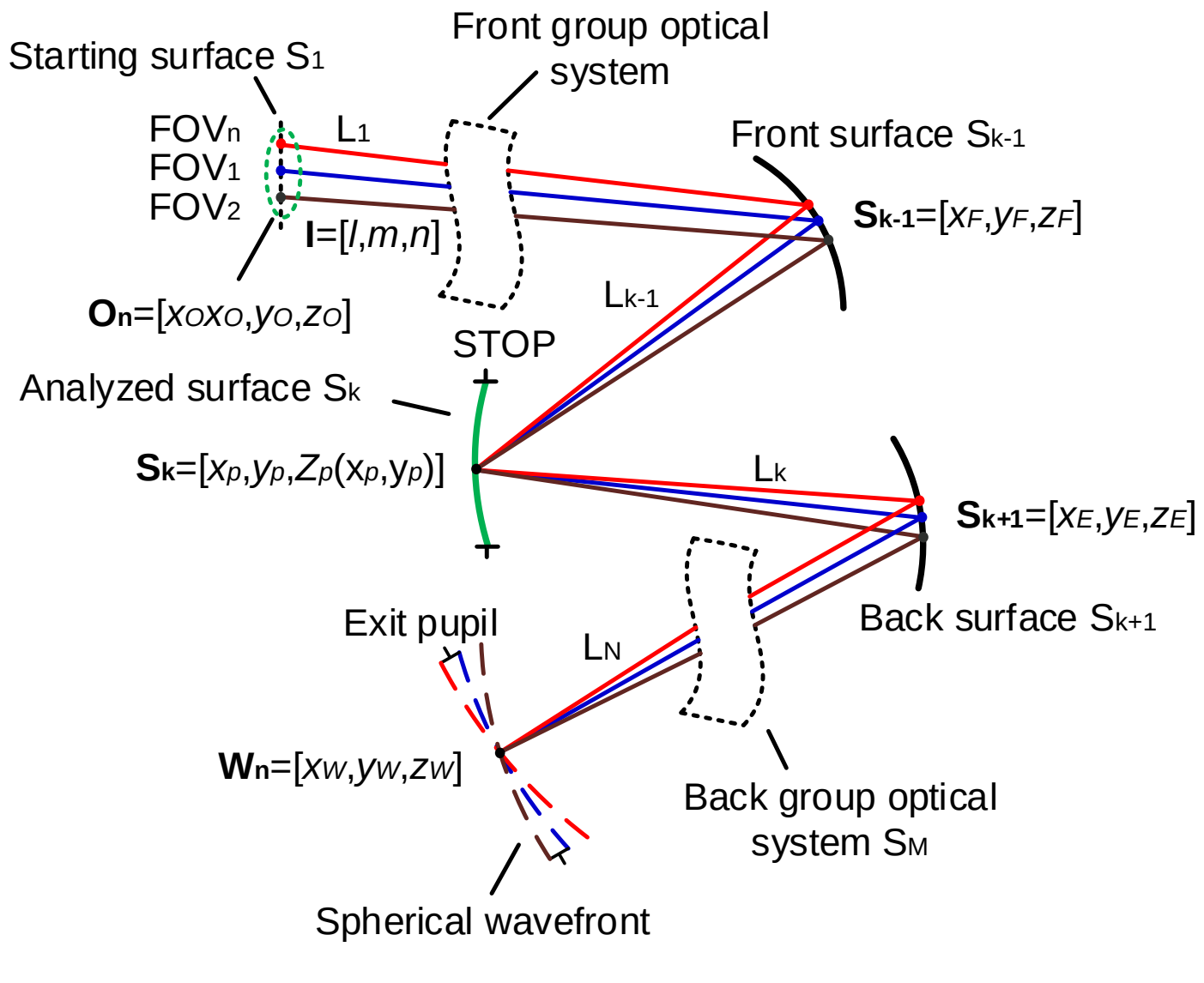

Fig. 6 Typical layout of a plane-symmetric freeform optical system

A direct approach is using real ray data to optimize Eq. (5), while adhering to the constraints outlined in Eq. (8), to obtain the surface error coefficient. However, multiple ray tracing during the optimization process diminishes computational efficiency. To enhance computation efficiency, a nested iterative approach was implemented. We maintain that the $x$ and $y$ coordinates of ray intersections on each surface, as well as the wavefront, remain constant throughout each iteration. When the shape of the optical surface is perturbed, the intersection points of the perturbed ray on the mirror no longer align with the reference spherical wavefront. As depicted in Fig. 7, **S** is no longer congruent with **S'**, and similarly, **W** no longer coincides with **W'**. The discrepancies in intersection points result in deviations from Eq. (15). Simultaneously, to mitigate changes in the ray position during each numerical handling process, two equality constraints are introduced to govern the propagation of the ray within the optical surface and the incident ray **I**. The equations are formulated as follows:

$$C_1 = \sqrt{\sum_{f=1}^{F}\sum_{j=1}^{R}((DL_x^{j,s})_f{}^2 + (DL_y^{j,s})_f{}^2)\Big/(F*R)}, \tag{16}$$

$$C_2 = \sum_{f=1}^{F}\sum_{j=1}^{R}\left(\left|(\mathbf{S}_2 - \mathbf{O}_n)\times\vec{\mathbf{I}}\right|\Big/(F*R)\right), \tag{17}$$

where × represents cross-product operation and $|\cdot|$ is modulo operations. The $\mathbf{S}_2$, $\mathbf{O}_n$ and **I** as illustrated in Fig. 6. Here, $DL_x^{j,s}$ and $DL_y^{j,s}$ can be derived by Eq. (15).

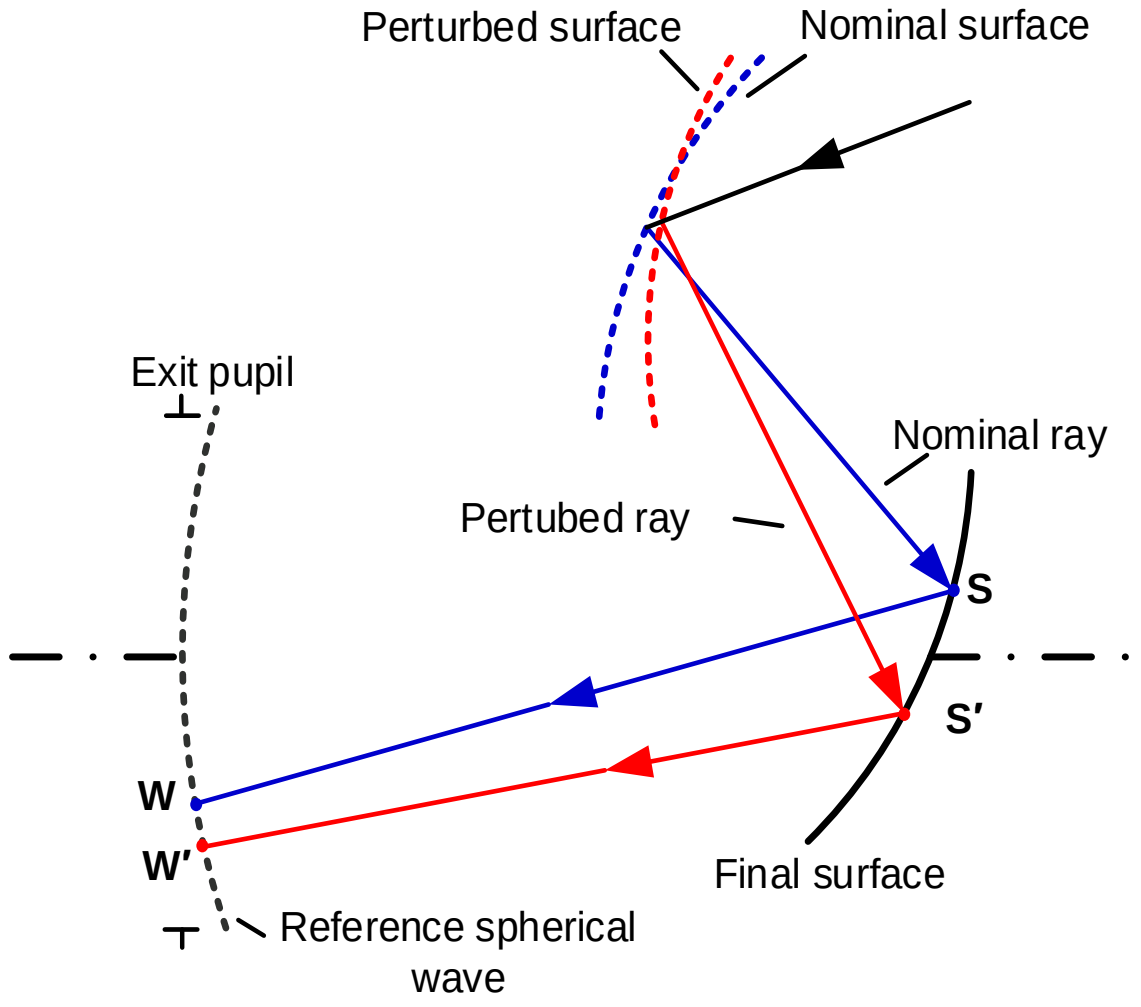

Fig. 7 Schematic diagram of error in ray operation

The resulting equations are nonlinear expression involving the surface error coefficient. The process of handling nonlinear equations with constraints is transformed into a numerical process, which is solved by minimizing the function $L(\boldsymbol{\alpha})$, expressed as follows:

$$\boldsymbol{\alpha}^* = \arg\min_{\boldsymbol{\alpha}} L(\boldsymbol{\alpha}), \tag{18}$$

$$L(\boldsymbol{\alpha}) = \omega_1 f_1(\boldsymbol{\alpha}) + \omega_2 C_1(\boldsymbol{\alpha}) + \omega_3 C_2(\boldsymbol{\alpha}), \tag{19}$$

where $\boldsymbol{\alpha}$ is the optimization variable, which can be the error surface coefficient. And $\omega_1$, $\omega_2$, and $\omega_3$ are the weights of the three parts of the function $L(\boldsymbol{\alpha})$. We scale the three components of function $L(\boldsymbol{\alpha})$ using weights to ensure that $f_1$, $C_1$, and $C_2$ are of similar orders of magnitude. The inequality constraint of the solution process is Eq. (8). The process can be dealt with quickly using the *fmincon* method in Matlab.

Once the numerical algorithm reaches the predetermined stopping criterion, the surface error coefficients resulting from this derive are incorporated into the optical surface. This is followed by the complete FOV ray tracing of the optical system, updating the *x* and *y* ray coordinates on the optical surface as well as the wavefront for the subsequent operation. Employing nested iterations, the approximation of the surface shape change is achieved while adhering to the efficiency requirement set by the wavefront boundary. The flowchart depicting the obtaining process of surface variation based on real ray data and surface frequency slopes is presented in Fig. 8.

In this paper, we employ Zernike polynomials [37] to describe the surface error. Therefore, the expression for the surface sag of the analyzed optical surface can be expressed as:

$$z_{\text{error}}(x, y, A_{nm}, \rho, \psi) = A_{nm} R_n^m(\rho) \cdot (\cos(m\psi) + i\sin(m\psi)), \tag{20}$$

$$R_n^m(\rho) = \sum_{k=0}^{\frac{n-m}{2}} \frac{(-1)^k \cdot (n-k)!}{k! \cdot (\frac{n+m}{2} - k)! \cdot (\frac{n-m}{2} - k)!} \cdot \rho^{n-2k}, \tag{21}$$

where $A_{nm}$, $n$, and $m$ are the coefficients of Zernike polynomial, polynomial order, and radial order, respectively. Here, $\rho$ and $\varphi$ represent angle and radial distance in polar coordinates. It is worth noting that the error equation in the algorithm described in this paper is also applicable to other surface functions, such as Gaussian radial basis function, *XY* polynomial, and so on.

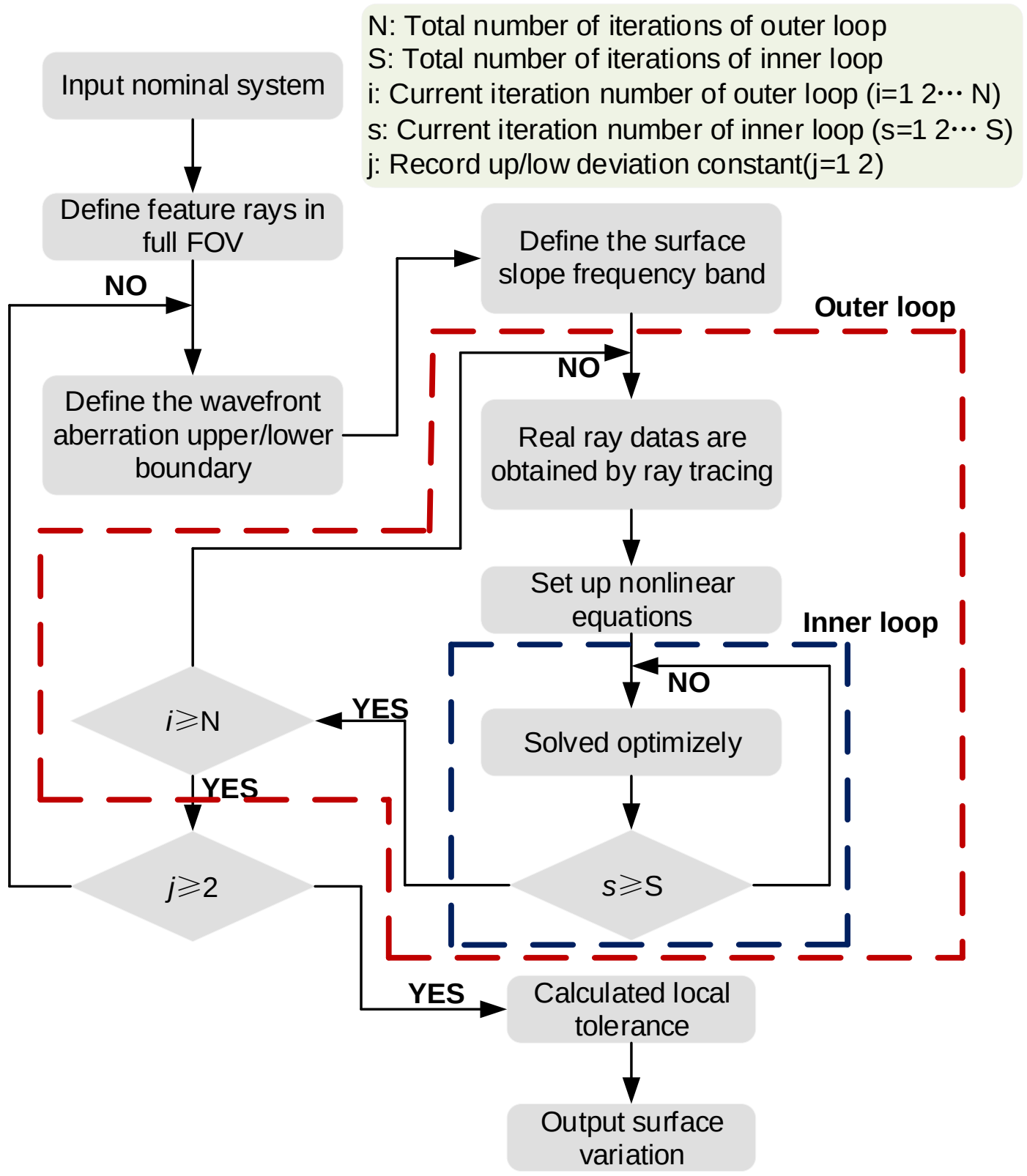

Fig. 8 Flowchart of the design process

The method presented in this paper leverages full FOV real ray tracing data, enabling the simultaneous obtaining of surface variation across the entire FOV. This approach is suited for freeform optical systems even in the presence of significant wavefront aberrations. Additionally, this method obviates the necessity of tracking the exit/entrance pupil position after imaging each optical surface or determining the wavefront formed by intermediate optical elements. It solely requires knowledge of the final exit pupil position, which can be readily obtained from optical software such as Zemax and CodeV.

## 3. Surface error analysis with two methods

In this section, we demonstrate the effectiveness of the proposed automated surface error analysis methodology for a single-field off-axis optical system consisting of a single freeform surface. This demonstration involves introducing wavefront disturbances of varying magnitudes and comparing the results between the method that considers changes in the normal field presented in this paper and the method [25] that disregards these changes. The operational wavelength of the optical system is 0.5μm, with a focal length of 179 mm and an aperture stop size of 60 mm. For reference, the optical path diagram and nominal wavefront aberration are illustrated in Fig. 9.

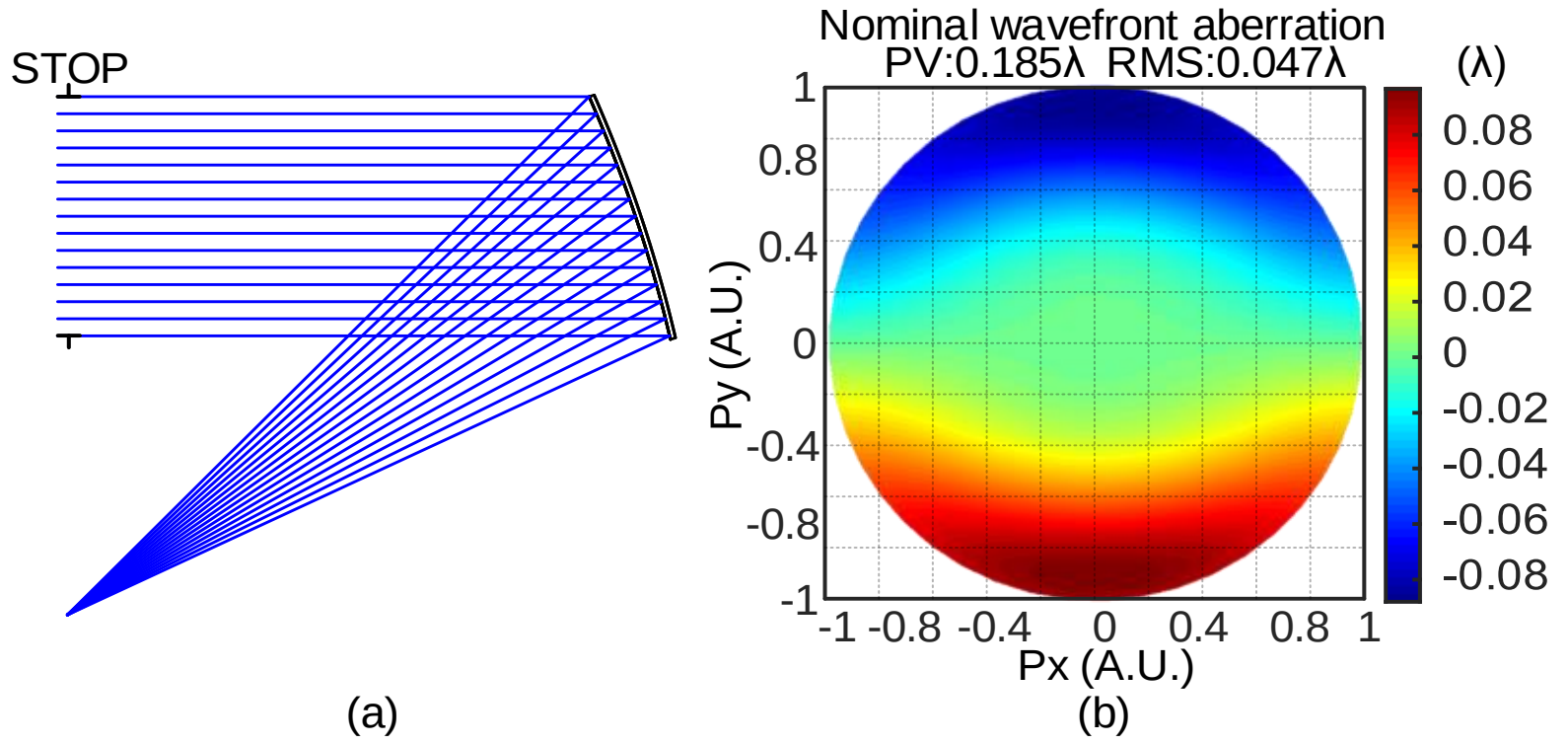

Fig. 9 Off-axis freeform surface single FOV optical system: (a) layout diagram; (b) nominal wavefront aberration

Next, we introduced wavefront disturbances with RMS values of 0.5 (RMS=0.025λ), 1 (RMS=0.05λ), 2.5 (RMS=0.125λ), and 25 (RMS=1.25λ) times the design value. We then combined each of these disturbances with the nominal wavefront to create the experimental wavefront, using two different methods to calculate the perturbations of the surface. The four sets of experimental wavefront diagrams are presented in Fig. 10.

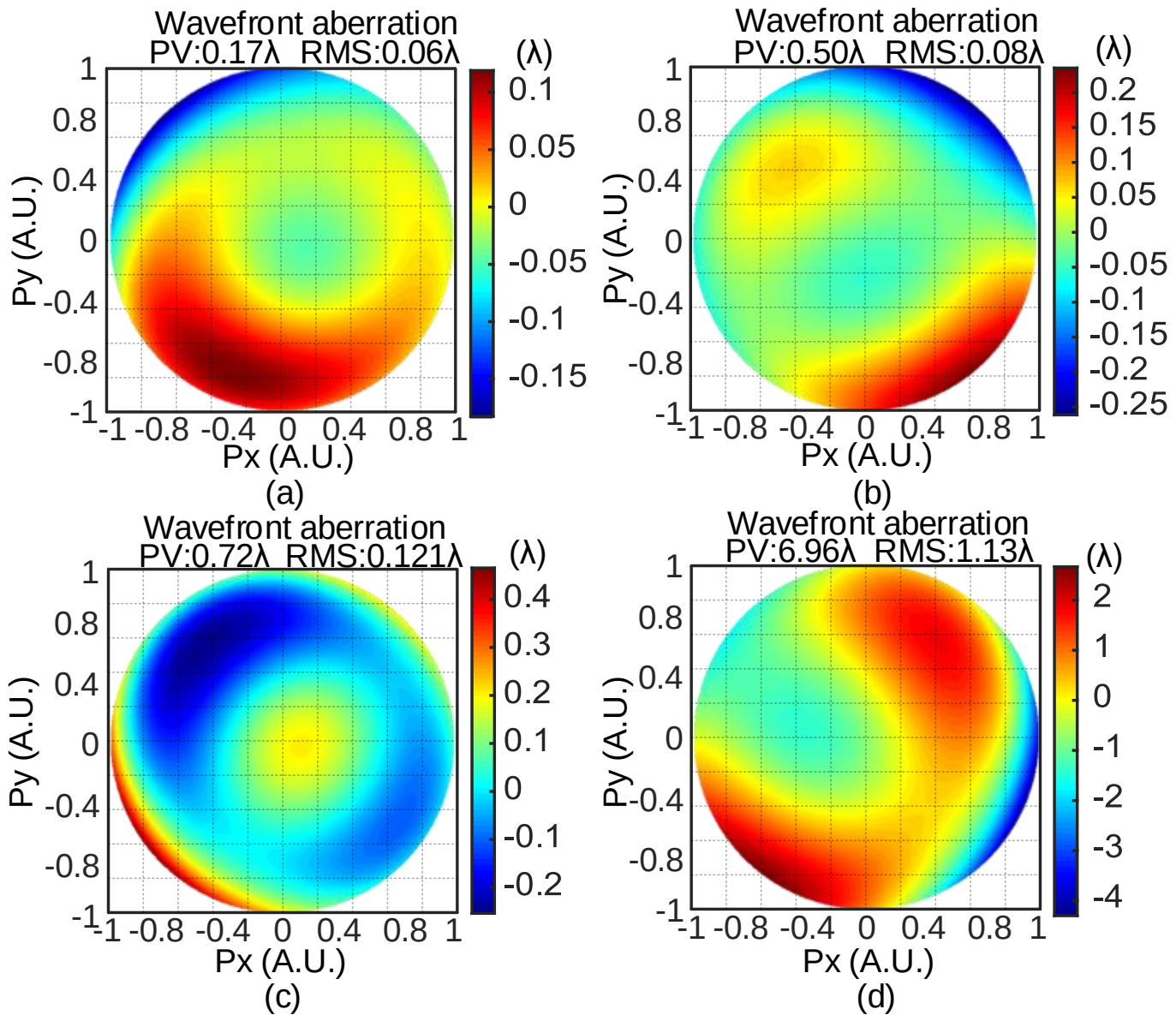

Fig. 10 Experimental wavefront aberration: (a) 0.5 times; (b) 1 times; (c) 2.5 times; (d) 25 times.

We applied the surface errors to the optical surface for forward ray tracing and computed the resulting wavefront aberrations, as shown in Fig. 11. Figures 11 (a)~(d) represent the forward tracing results of the method that takes changes in the normal field into account, while Figs. 11 (e)~(h) depicts the forward tracing results of the method that disregards changes in the normal field. The surface errors obtained by the two methods are presented in Figs. 12(a)~(d) and Figs. 12 (e)~(h), respectively. Both methods performed well in recovering the expected wavefront aberration for small surface sag errors. However, as the surface error increased, the method that considered changes in the normal field continued to recover the anticipated wavefront aberrations accurately. Conversely, the method that neglected alterations in the normal vectors produced inaccuracies in the reconstructed wavefront aberration due to surface normal errors.

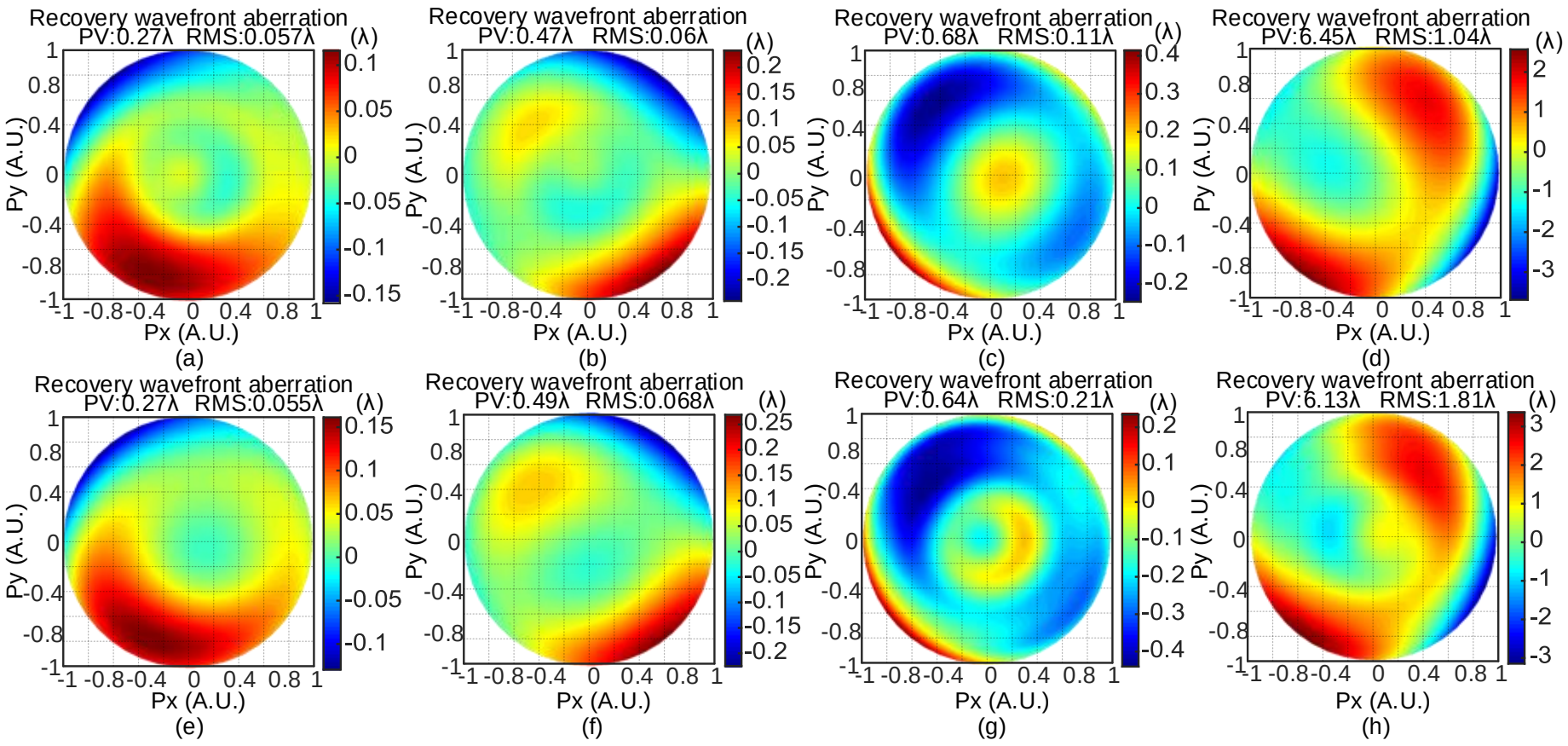

Fig. 11 Wavefront aberration by forward tracing: (a)~(d) method that consider changes in normals; (e)~(h) method that ignoring changes in normals

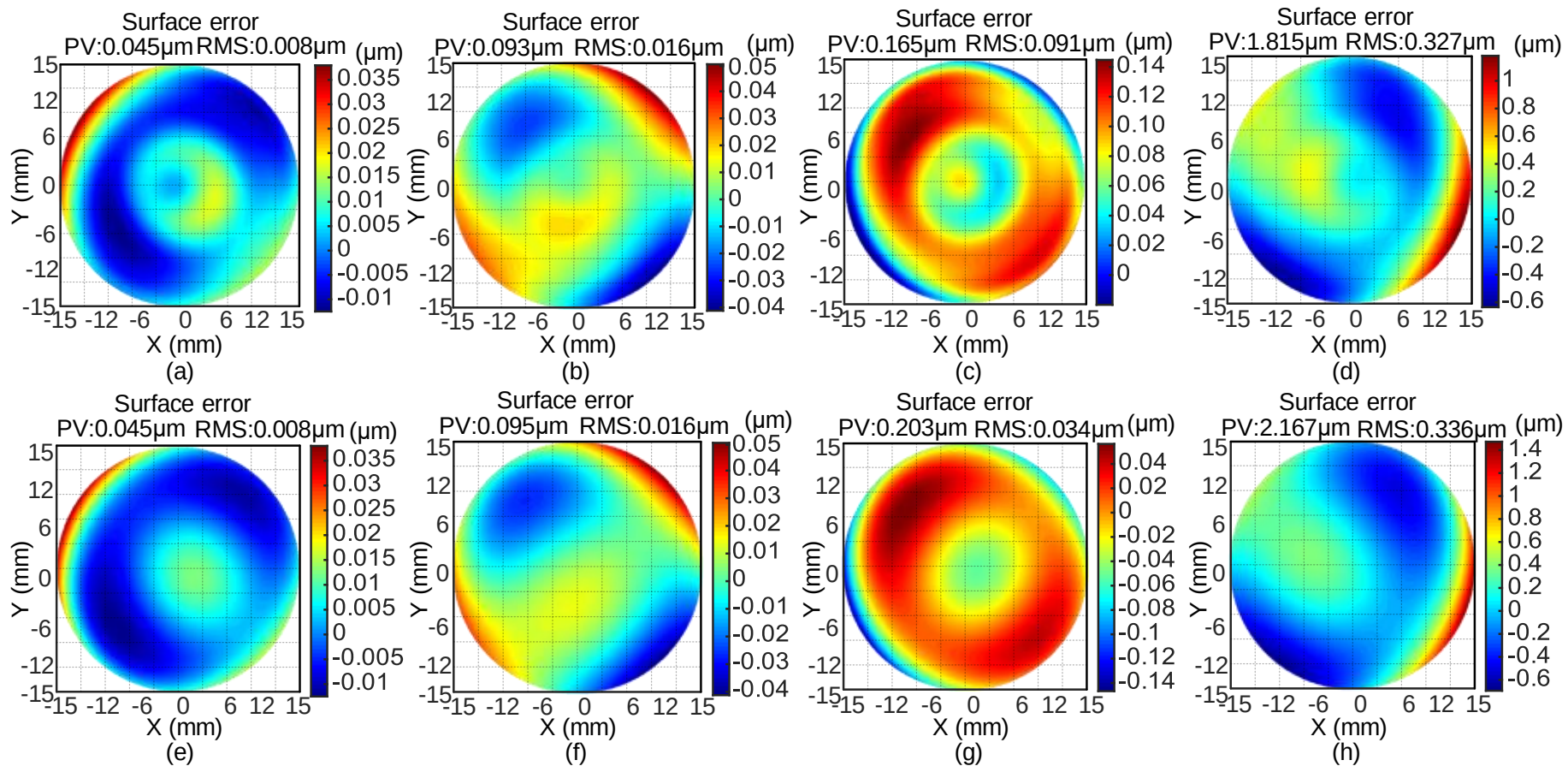

Fig. 12 Surface error obtained by two methods: (a)~(d) method that consider changes in normal field; (e)~(h) method that ignoring changes in normal field.

## 4. Surface variation analysis of different frequency bands

To demonstrate the impact of surface slope at different frequency segments on the sag variation values of the surface, we will employ a two-mirror off-axis freeform reflective system in this section, which operates in the near-infrared band ($\lambda$=1μm). It has a focal length of 416.5mm, a F-number of 5, and a FOV of 1°×1°. Reflective optical systems, being free from chromatic aberration, do not factor in the influence of chromatic aberrations on the surface error of freeform surfaces. This system is symmetrical about the YOZ plane in which the aperture stop is positioned in front of the first mirror. All mirrors are freeform surfaces that can be described using *XY* polynomials. The system layout is depicted in Fig. 13(a), while the full FOV map of wavefront aberrations is presented in Fig. 13(b). In Fig. 13(b), the average value is 0.045λ, the PV maximum is 0.5λ, and the RMS maximum is 0.052λ. Our proposed methodology is applied to calculate surface sag variation of the second mirrors across various frequency bands. The FOV was sampled in a uniform manner using a rectangular grid, resulting in a total of 31 field samples. The ray data for a single field was sampled radially, yielding a total of 45×45 samples for each single field.

The maximum wavefront PV value of the system is 0.5λ. To achieve acceptable imaging performance for most fields, the wavefront PV threshold of 0.6λ is employed during the operation. The widely employed 37 term Zernike polynomial provides the characterization of low to intermediate-frequency errors on optical surfaces. To broaden the analyzed frequency range, we employ 70 term Zernike coefficients for tolerance assessment. Typically, a spatial frequency below 10 is considered intermediate [34]. Therefore, we analyzed surface tolerance within the 0~8 frequency band. Additionally, this approach can be extended to higher frequencies using alternate expressions. Then, the sag variation values of the second mirrors are operated within the frequency bands of 0~8, and the analysis results of SlopeRMS and surface variation are shown in Fig. 14. We selected three frequency bands from the onset and conclusion of the analysis results, as well as the interior, for subsequent analysis. The details of surface variation in the three frequency bands of 0~0.5, 1~1.5 and 6~8 are shown in Fig. 15. The upper and lower variation are showcased in Figs. 15(a) and (b), respectively. Figures. 15(c) shows the surface variation map, which is the difference between the upper and the lower variation.

As the frequency band widens, we demonstrate an increased variation in the surface slope. Simultaneously, as the optical surface variation becomes steeper, the overall permissible surface variation RMS value for surface sag tightens. Additionally, as depicted in Fig. 15, the upper and lower boundaries for surface variation are no longer consistently positive or negative. This phenomenon can be attributed to the intricate influence of the surface normal field on rays, resulting in interdependent point variation. Consequently, it implies that when controlling the surface normal field within a specific range of variation, certain surface regions may not allow for surface depressions or protrusions. Regarding the surface variation in the three frequency bands, the loosest values are 0.42μm, 0.41μm, and 0.12μm, while the tightest values are 0.23μm, 0.043μm, and 0.014μm.

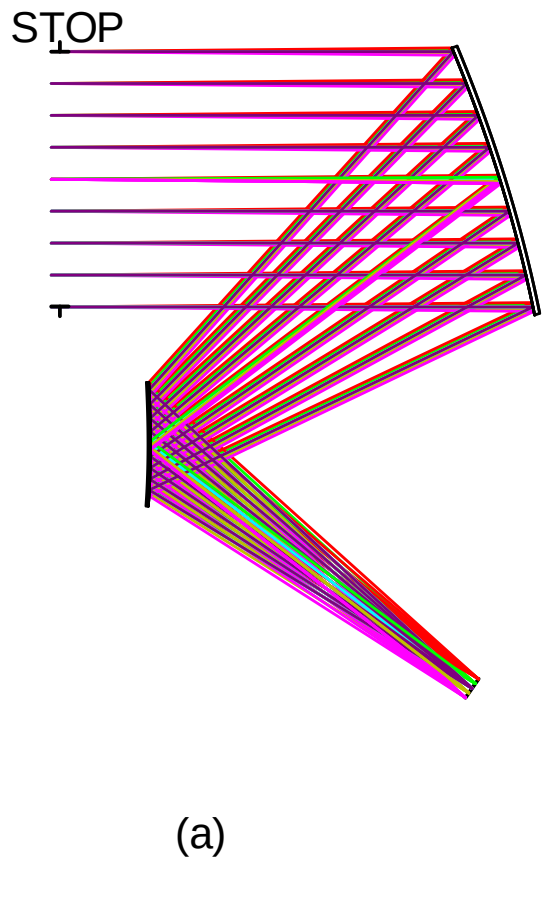

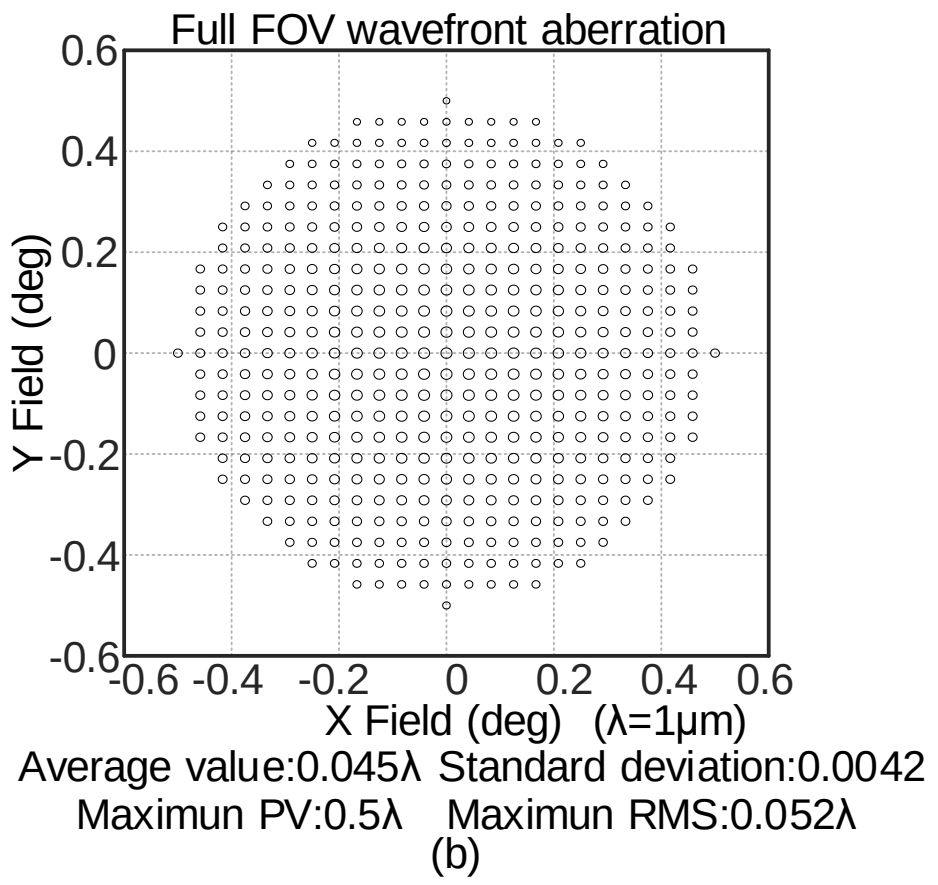

Fig. 13 (a) Two-mirror off-axis reflective system layout; (b) Field map of the rms wavefront aberration for the two-mirror off-axis reflective system

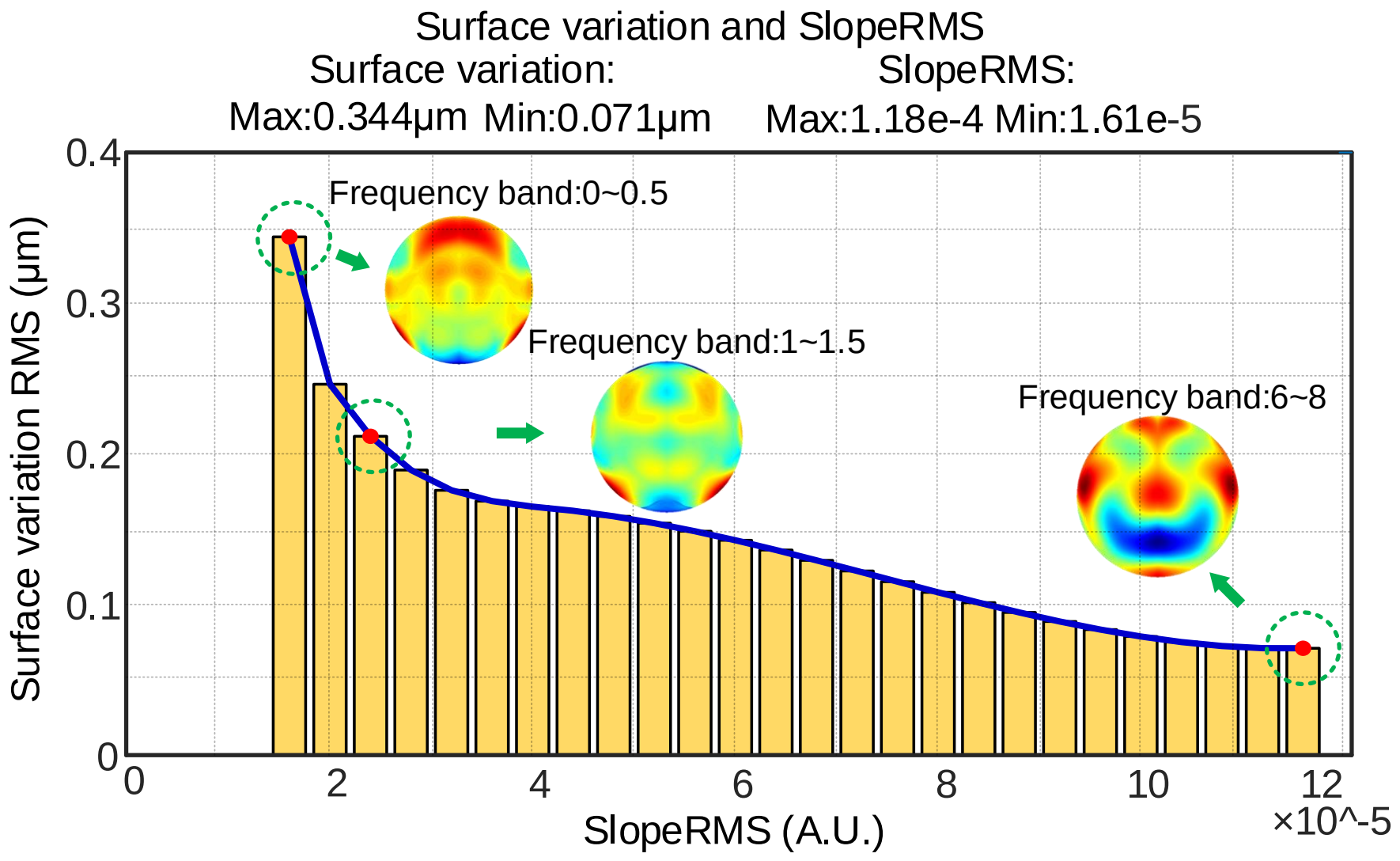

Fig. 14. Function diagram between SlopeRMS and surface variation

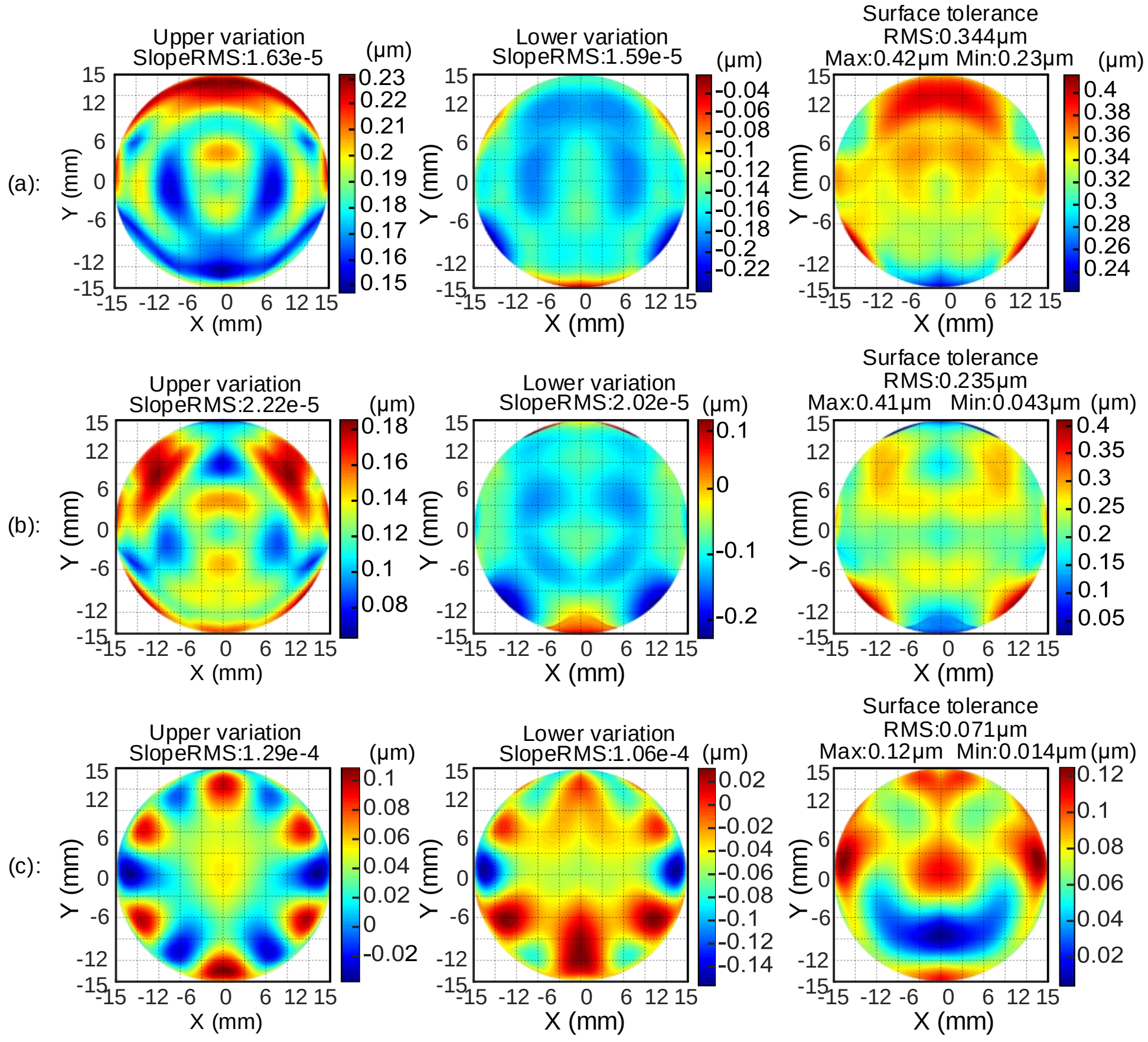

Fig. 15. Upper, lower variation boundary and surface variation: (a) frequency band of 0~0.5; (b) frequency band of 1~1.5; (c) frequency band of 6~8.

In this paper, computations were implemented in MATLAB 2018b on a computer with an Intel i7-10700K CPU and 16GB (RAM), running Windows 10. Due to the simultaneous processing of the full FOV data and the absence of complex ray tracing in the process of numerical solution, obtaining this example surface variation of each frequency band took around 30 minutes.

To ensure consistency in the frequency bands for surface error generation and to verify surface tolerance across the three frequency bands, we conducted Monte Carlo experiments by randomly generating 2000 surface errors using 70 terms of standard Zernike polynomials for each frequency band and subsequently applied them to the second mirror. Subsequently, observing the PV value of the wavefront aberration across the system. In each Monte Carlo test group, we analyzed the PV values of wavefront aberrations for 31 sampling fields. The Monte-Carlo analysis results for the three frequency bands are illustrated in Fig. 16, where the PV values of wavefront aberration are lower than the preset threshold in 90% of the experimental groups across the full FOV for all three Monte-Carlo analysis sets. Notably, no instances of excessively good or excessively poor experimental results were observed, which could be attributed to the both influence of surface sag and slope variation. These results underscore the effectiveness of the surface variation model based on different frequency bands.

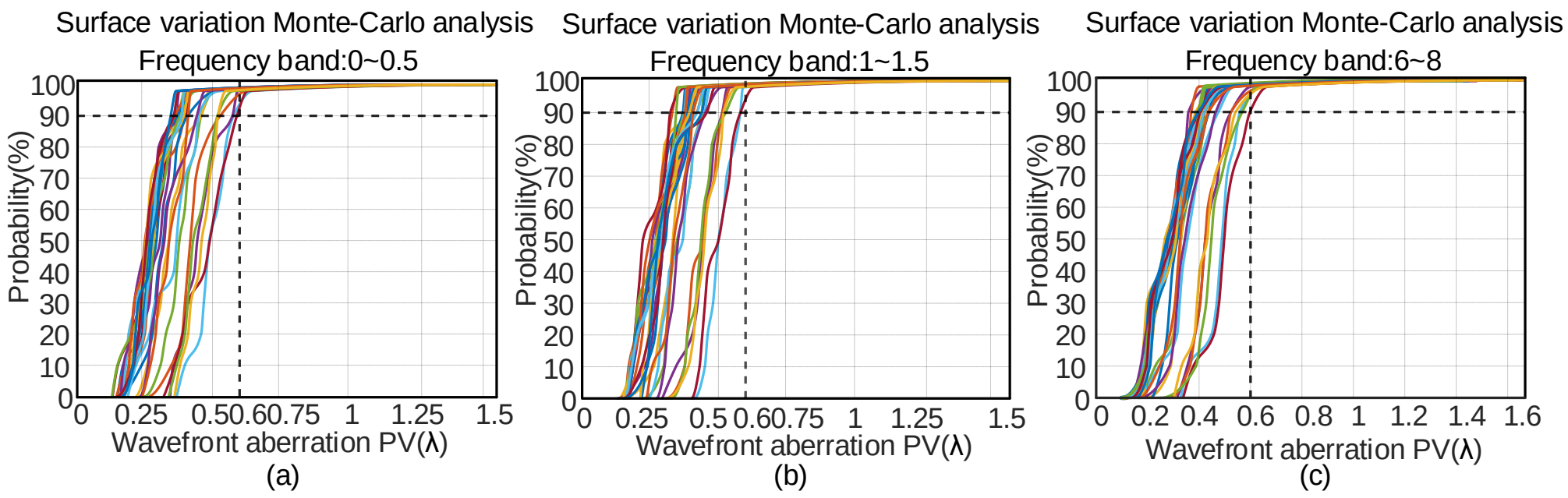

Fig. 16 Monte-Carlo analysis for surface variation with three different frequency band: (a) frequency band of 0~0.5; (b) frequency band of 1~1.5; (c) frequency band of 6~8.

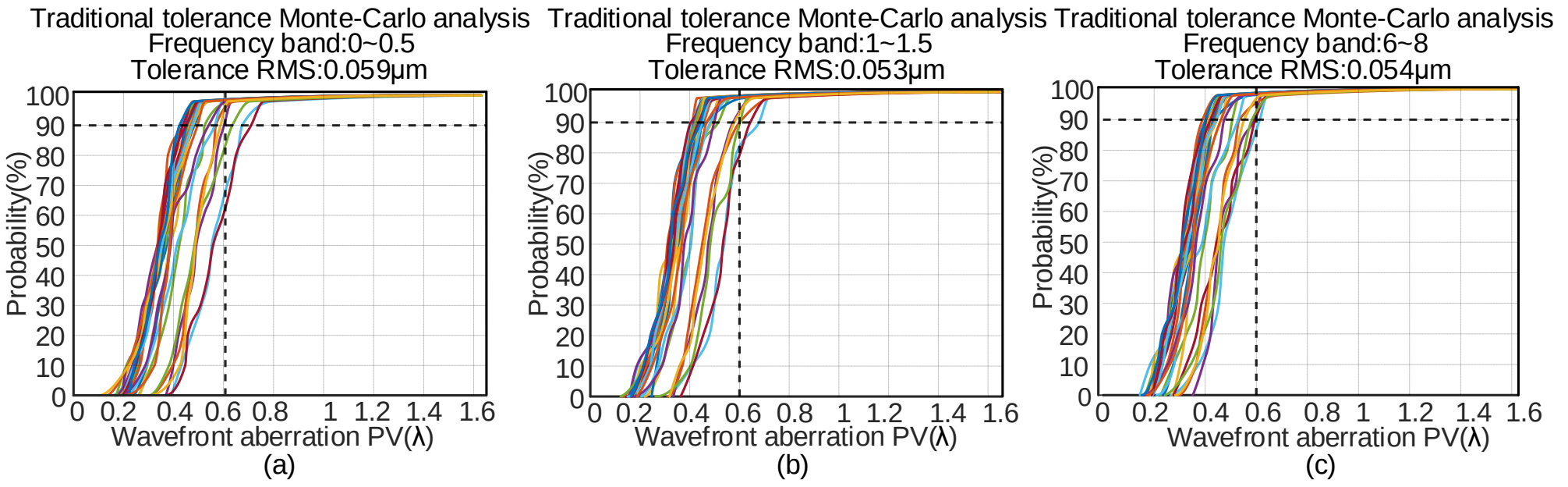

Fig. 17. Monte-Carlo analysis for surface tolerance with traditional model: (a) frequency band of 0~0.5; (b) frequency band of 1~1.5; (c) frequency band of 6~8.

The surface variation method described herein efficiently assigns the surface variation to each sampled region, resulting in more forgiving surface variation compared to traditional tolerance analysis methodologies. Particularly, the RMS value of the surface variation derived by this method surpasses that of the traditional tolerance analysis. This distinction is illustrated in Fig. 17, which presents the outcomes of Monte Carlo experiments conducted using the traditional tolerance analysis. In this context, the tolerance RMS for the three distinct frequency bands recorded as 0.059μm, 0.053μm, and 0.054μm, with an average value of 0.056μm. The traditional tolerance model does not differentiate between frequency bands. Thus, the tolerance analysis for different frequency bands is expected to be similar, and the results also validate this observation.

## 5. Conclusion

In this study, we introduce an innovative approach that utilizes real ray data and surface slopes at different frequencies to analyze surface variation. The approach, which includes establishing nonlinear equations based on prescribed wavefront errors, enhances computational speed and accuracy through nested iterations. Additionally, this method accounts for changes in ray direction resulting from variation in the surface slope. By taking the single-field single-mirror system and the off-axis two-mirror system as examples, we demonstrate the potential of the method from many perspectives, including high efficiency, capability to tackle systems with relatively large wavefront aberration, and the ability to analyze surface variation with different frequency band slope. Simultaneously, we find that a negative correlation between surface slope fields and the RMS value of surface variation has been demonstrated. Furthermore, when controlling the surface normal field within a specific range of variation, it restricts surface depressions or bulges in specific regions. We anticipate that the phenomenon will expand the horizons of surface variation analysis in freeform optical systems designed for various applications.

Our forthcoming research endeavors will concentrate on extending this algorithm to simultaneously obtain surface variation for multiple freeform surfaces, which can pave the way for future freeform surface desensitization design and manufacturing. Furthermore, the method can be expanded to refractive freeform optical elements.

### Disclosures

The authors declare no conflicts of interest.

### Funding

National Key Research and Development Program of China (2023YFC2414700); National Natural Science Foundation of China (12274156); Science, Technology and Innovation Commission of Shenzhen Municipality (JCYJ20210324115812035).

### Data availability

Data underlying the results presented in this paper are not publicly available at this time but may be obtained from the authors upon reasonable request.